# Decay Mode Solutions to Cylindrical KP Equation

Mingliang Wang[1,2], Jinliang Zhang[1*] & Xiangzheng Li[1]

1. School of Mathematics & Statistcs, Henan University of Science & Technology, Luoyang, 471023, PR China

2. School of Mathematics & Statistcs, Lanzhou University, Lanzhou, 730000, PR China

* Corresponding author. E-mail: zhangjin6602@163.com

**Abstract:** A nonlinear transformation for the cylindrical KP(CKP) equation has been derived by using the simplified homogeneous balance method (SHB). The 1-decay mode and 2-decay mode solutions of the CKP equation have been obtained in terms of the nonlinear transformation derived here. The results obtained in the paper are different from those obtained in the previous literaturs.

**AMS(2000) Subject Classification:** 35Q20; 35Q53

## 1. Introduction

In the present paper, we will investigate the cylindrical KP equation (CKP) in the form

$$\left(u_t + 6uu_x + u_{xxx}\right)_x + \frac{1}{2t}u_x + \frac{3\alpha^2}{t^2}u_{yy} = 0. \tag{1}$$

which was introduced by Johnson[1-2] to describe surface wave in a shallow incompressible fluid. The CKP Eq.(1) for magnetized plasmas with pressure effects and transverse perturbations in cylindrical geometry was also derived by using the small amplitude perturbation expansion method[3]. And Eq.(1) is a (2+1)-dimensional generalization of the cylindrical KdV(CKdV) equation[4-5]

$$u_t + 6uu_x + u_{xxx} + \frac{1}{2t}u = 0. \tag{2}$$

Due to the importance and wide application , CKP Eq.(1) has been paid attention by many researchers in mathematical physics. For instance, In Ref.[6], Klein et al have shown that the Lax pair corresponding KP and CKP equation are gauge equivalent, some class of solutions ( such as horseshoelike-front solutions, lump solutions and rational solutions ) were obtained by using Darboux transformation approach. In Ref.[7], Deng has shown that the decay mode solution for CKP Eq.(1) can be obtained by Backlund transformation and Hirota's method.

In this paper, we will use the simplified homogeneous balance method(SHB)[8-12] to derive a nonlinear transformation for the CKP equation, based on the transformation derived here, the 1-decay mode and 2-decay mode solutions of CKP equation can be obtained. These results are different form those obtained in the previous literatures.

## 2. Derivation of the nonlinear transformation

Considering the homogeneous balance between $uu_x$ and $u_{xxx}$ in Eq.(1)

$(2m+1=m+3 \to m=2)$, according to the SHB[8], which means that an undetermined function $F(\phi)$ and its partial derivatives $F_x = F'\phi_x$ ..., that appearing in the original homogeneous balance (BH) [9-12] are replaced by a logarithmic function $A\ln(\phi)$ and its derivatives $A(\ln\phi)_x = A\frac{\phi_x}{\phi},...$, respectively, we can suppose that the solution of Eq.(1) is of the form

$$u(x,y,t) = A(\ln\varphi)_{xx} + \frac{x+\lambda y}{12t}, \qquad (3)$$

where the constant $A$ and the function $\varphi = \varphi(x,y,t)$ are to be determined, $\frac{x+\lambda y}{12t} \equiv u_0$ be a particular solution of Eq.(1), $\lambda$ – arbitrary constant. we aim to determine constant A and the function $\varphi = \varphi(x,y,t)$ such that the expression (3) satisfies Eq.(1) exactly.

Substituting (3) into the left hand side of Eq.(1), noticing that

$$(u_{0t} + 6u_0 u_{0x} + u_{0xxx})_x + \frac{1}{2t} u_{0x} + \frac{3\alpha^2}{t^2} u_{0yy} = 0,$$

yields

$$(u_t + 6uu_x + u_{xxx})_x + \frac{1}{2t} u_x + \frac{3\alpha^2}{t^2} u_{yy}$$

$$= A\frac{\partial^2}{\partial x^2}\left[ (\ln\varphi)_{xt} + 3A(\ln\varphi)_{xx}^2 + 6u_0(\ln\varphi)_{xx} \right.$$

$$\left. + (\ln\varphi)_{xxxx} + \frac{1}{2t}(\ln\varphi)_x + \frac{3\alpha^2}{t^2}(\ln\varphi)_{yy} \right]$$

$$= A\frac{\partial^2}{\partial x^2}\left[ \frac{(\varphi_t + u_0\varphi_x + \varphi_{xxx})_x + \frac{3\alpha^2}{t^2}\varphi_{yy}}{\varphi} \right.$$

$$+ \frac{-\varphi_t\varphi_x + (3-3A)\varphi_{xx}^2 + 4\varphi_x\varphi_{xxx} + 6u_0\varphi_x^2 - \frac{3\alpha^2}{t^2}\varphi_y^2}{\varphi^2} \qquad (4)$$

$$\left. + \frac{(12-6A)\varphi_x^2\varphi_{xx}}{\varphi^3} + \frac{(3A-6)\varphi_x^4}{\varphi^4} \right].$$

Setting the coefficient of $\varphi^{-4}$ to zero, yields $A=2$, by using this, the expression (3) becomes

$$u(x,y,t) = 2(\ln\varphi)_{xx} + \frac{x+\lambda y}{12t} \qquad (5)$$

and the expression (4) can be simplified as

$$(u_t + 6uu_x + u_{xxx})_x + \frac{1}{2t}u_x + \frac{3\alpha^2}{t^2}u_{yy}$$

$$= 2\frac{\partial^2}{\partial x^2}\left\{\frac{1}{\varphi^2}\left[\varphi\left(\left(\varphi_t + \varphi_{xxx} + \frac{x+\lambda y}{12t}\varphi_x\right)_x + \frac{3\alpha^2}{t^2}\varphi_{yy}\right)\right.\right.$$

$$\left.\left.-\left(\varphi_t\varphi_x + 4\varphi_x\varphi_{xxx} - 3\varphi_{xx}^2 + \frac{x+\lambda y}{2t}\varphi_x^2 + \frac{3\alpha^2}{t^2}\varphi_y^2\right)\right]\right\} = 0 \quad (6)$$

provided that the function $\varphi = \varphi(x, y, t)$ satisfies the homogemeity equation

$$\varphi\left[\left(\varphi_t + \varphi_{xxx} + \frac{x+\lambda y}{12t}\varphi_x\right)_x + \frac{3\alpha^2}{t^2}\varphi_{yy}\right]$$

$$-\left(\varphi_t\varphi_x + 4\varphi_x\varphi_{xxx} - 3\varphi_{xx}^2 + \frac{x+\lambda y}{2t}\varphi_x^2 + \frac{3\alpha^2}{t^2}\varphi_y^2\right) = 0. \quad (7)$$

From (5) (6) and (7), we come to the conclusion that if $\varphi = \varphi(x, y, t)$ is a solution of the homogemeity equation (7), substituting it into (5), we have the exact solution of the CKP equation (1). The expression (5) with the homogemeity equation (7) together is called the nonlinear transformation for the CKP equation.

In the next sections, using the nonlinear transformation derived here, we will find the exact solutions of the CKP equation (1).

## 3. 1-Decay mode solution

In view of the homogemeity of Eq.(7) we suppose that the solution of Eq.(7) is of the form

$$\varphi(x, y, t) = 1 + e^\eta, \eta = B(t)x - \int_0^t \left[B^3(t) + \frac{3\alpha^2\lambda^2 B(t)}{t^2}\right]dt \quad (8)$$

where $B(t)$ to be determined later.

Substituting (8) into the left hand side of Eq.(7), yields

$$\left[(1 + B(x + \lambda y))(B' + \tfrac{1}{2t}B)\right]e^\eta + \left[B' + \tfrac{1}{2t}B\right]e^{2\eta} = 0,$$

provided that $B(t)$ satisfies the ODE

$$B' + \tfrac{1}{2t}B = 0. \quad (9)$$

Solving ODE (9), we have

$$B(t) = \frac{k}{\sqrt{t}}, \quad (10)$$

where $k$ be an arbitrary constant.

Substituting (10) into (8), we have a solution of Eq.(7)

$$\varphi = 1 + e^{\eta}, \eta = \frac{k}{\sqrt{t}}(x + \lambda y) + \frac{2k^3}{\sqrt{t}} + \frac{2\alpha^2 \lambda^2 k}{t^{3/2}} + c, \quad (11)$$

Substituting (11) into the expression (5), we have an exact 1-decay mode solution of the CKP equation (1), as follows

$$u(x, y, t) = \frac{k^2}{2t} \operatorname{sech}^2\left(\tfrac{1}{2}\eta\right) + \frac{x + \lambda y}{12t}, \quad (12)$$

where $\eta$ is expressed by (11).

The solution (12) includes two parts : One is a bell type of decay wave $u_1 = \frac{k^2}{2t} \operatorname{sech}^2\left(\tfrac{1}{2}\eta\right)$ when the variable $t$ is increasing, which is not a solution of Eq.(1); The other one is the paticular solution $u_0 = \frac{x + \lambda y}{12t}$ of Eq.(1).

If taking $\alpha = 0$, then (12) becomes

$$u(x, y, t) = \frac{k^2}{2t} \operatorname{sech}^2 \frac{1}{2}\left(\frac{k}{\sqrt{t}}(x + \lambda y) + \frac{2k^3}{\sqrt{t}}\right) + \frac{x + \lambda y}{12t}, \quad (13)$$

which is the solution of CKdV equation (2), where $\lambda y$ can be thought of as an arbitrary parameter.

The solutions (12) and (13) are the 1-decay mode solutions of CKP equation and CKdV equation respectively, when $t$ is increasing; and these results are different form that obtained in [1]- [7], where the solutions of the CKP equation and CKdV equation were expressed in terms of the Airy functions.

## 4. 2-decay mode solution

In order to find more general solutions of the CKP equation (1), we need find more general solutions of the homogeneity equation (7). We suppose that the solution of Eq.(7) is of the form

$$\phi(x, y, t) = 1 + \varepsilon \phi^{(1)} + \varepsilon^2 \phi^{(2)} + \varepsilon^3 \phi^{(3)} + \ldots \quad (14)$$

where $\phi^{(i)} (i = 1, 2, 3, \ldots)$ to be determined later, $\varepsilon$ be a small parameter (we may take $\varepsilon = 1$ for simplicity). Substituting (14) into (7), collecting all terms with $\varepsilon^i$ together($i = 1,2,3\ldots$), and setting the coefficient of $\varepsilon^i$ to zero (i=1,2,3,…), yields a hierarchy of equations for $\phi^{(i)} (i = 1, 2, 3, \ldots)$ as follows

$$\varepsilon: \quad \left(\phi_t^{(1)} + \phi_{xxx}^{(1)} + \tfrac{x+\lambda y}{2t}\phi_x^{(1)}\right)_x + \frac{3\alpha^2}{t^2}\phi_{yy}^{(1)} = 0, \tag{15}_1$$

$$\varepsilon^2: \quad \begin{aligned} &\left(\phi_t^{(2)} + \phi_{xxx}^{(2)} + \tfrac{x+\lambda y}{2t}\phi_x^{(2)}\right)_x + \frac{3\alpha^2}{t^2}\phi_{yy}^{(2)} = \\ &\left[\phi_x^{(1)}\phi_t^{(1)} + 4\phi_x^{(1)}\phi_{xxx}^{(1)} - 3(\phi_{xx}^{(1)})^2 + \tfrac{x+\lambda y}{2t}(\phi_x^{(1)})^2 + \frac{3\alpha^2}{t^2}(\phi_y^{(1)})^2\right] \end{aligned} \tag{15}_2$$

$$\varepsilon^3: \quad \left(\phi_t^{(3)} + \phi_{xxx}^{(3)} + \tfrac{x+\lambda y}{2t}\phi_x^{(3)}\right)_x + \frac{3\alpha^2}{t^2}\phi_{yy}^{(3)} = -\phi^{(1)}\left[\left(\phi_t^{(2)} + \phi_{xxx}^{(2)} + \tfrac{x+\lambda y}{2t}\phi_x^{(2)}\right)_x + \frac{3\alpha^2}{t^2}\phi_{yy}^{(3)}\right]$$

$$+\phi_x^{(1)}\phi_t^{(2)} + \phi_x^{(2)}\phi_t^{(1)} + 4\phi_x^{(1)}\phi_{xxx}^{(2)} + 4\phi_x^{(2)}\phi_{xxx}^{(1)} - 6\phi_{xx}^{(1)}\phi_{xx}^{(2)}$$

$$+ \tfrac{x+\lambda y}{2t}(2\phi_x^{(1)}\phi_x^{(2)}) + \frac{3\alpha^2}{t^2}(2\phi_y^{(1)}\phi_y^{(2)}), \tag{15}_3$$

… ….

and so on, to be solved.

It should be noted that the hierarchy of Eq. $(15)_n$ ( $n = 1, 2, 3, ...$ ) are all linear equations, and the right hand side of any latter one depends only upon the former ones, so we can get $\phi^{(i)} = \phi^{(i)}(x, y, t)$ successively ( $i = 2, 3, ...$ ) by solving the same kind of linear equations provided that $\phi^{(1)} = \phi^{(1)}(x, y, t)$ is given. It is easy to give a solution of Eq. $(15)_1$ that is a linear one. As an illustrative example, we take

$$\phi^{(1)} = e^{\eta_1} + e^{\eta_2}, \quad \eta_i = A_i(t)(x+\lambda y) - \int_0^t\left[A_i^3 + \frac{3\alpha^2\lambda^2 A_i}{t^2}\right]dt, i=1,2. \tag{16}$$

where $A_i = \dfrac{k_i}{\sqrt{t}}$, $\eta_i = \dfrac{k_i}{\sqrt{t}}(x+\lambda y) + \dfrac{2k^3}{\sqrt{t}} + \dfrac{2\alpha^2\lambda^2}{t^{3/2}} + c, i=1,2.$

It is obvious that $\phi^{(1)}$ in (16) satisfies the linear Eq. $(15)_1$. Substituting (16) into the right hand side of Eq. $(15)_2$, yields

$$\left(\phi_t^{(2)} + \phi_{xxx}^{(2)} + \tfrac{x+\lambda y}{2t}\phi_x^{(2)}\right)_x + \frac{3\alpha^2}{t^2}\phi_{yy}^{(2)} = 3A_1A_2(A_1-A_2)^2 e^{\eta_1+\eta_2}, \tag{17}$$

which admits a solution

$$\phi^{(2)} = a_{12}e^{\eta_1+\eta_2}, \quad a_{12} = \frac{(A_1-A_2)^2}{(A_1+A_2)^2} = \frac{(k_1-k_2)^2}{(k_1+k_2)^2}. \tag{18}$$

Substituting (16), (17) and (18) into the right hand side of Eq. $(15)_3$, which vanishes, we

have

$$\left(\phi_t^{(3)} + \phi_{xxx}^{(3)} + \frac{x+\lambda y}{2t}\phi_x^{(3)}\right)_x + \frac{3\alpha^2}{t^2}\phi_{yy}^{(3)} = 0.$$

We may take $\phi^{(3)} = 0$ for its solution, hence $\phi^{(n)} = 0 (n \geq 3)$, so the series (14) truncates.

Substituting $\phi^{(1)}$ in (16) and $\phi^{(2)}$ in (18) as well as $\phi^{(n)} = 0 (n \geq 3)$ into (14) (take $\varepsilon = 1$), yields an exact solution of Eq.(7) as follows

$$\phi(x,y,t) = 1 + e^{\eta_1} + e^{\eta_2} + a_{12} e^{\eta_1+\eta_2}. \tag{19}$$

Substituting (19) into (5) we have the exact 2-decay mode solution of CKP equation as follows

$$u(x,y,t) = \frac{2}{t} \frac{k_1^2 e^{\eta_1} + k_2^2 e^{\eta_2} + (k_1-k_2)^2 e^{\eta_1+\eta_2} + a_{12}\left(k_2^2 e^{2\eta_1+\eta_2} + k_1^2 e^{\eta_1+2\eta_2}\right)}{\left[1 + e^{\eta_1} + e^{\eta_2} + a_{12} e^{\eta_1+\eta_2}\right]^2} + \frac{x+\lambda y}{12t} \tag{20}$$

where $\eta_i = \frac{k_i}{\sqrt{t}}(x+\lambda y) + \frac{2k_i^3}{\sqrt{t}} + \frac{2\alpha^2\lambda^2 k_i}{t^{3/2}}, i=1,2.$

The solution (20) of CKP equation includes two parts:
One is the big fractional expression;

The other part is $u_0 = \frac{x+\lambda y}{12t}$ which is a particular solution of CKP equation. The big fractional expression represents the interaction of two decay mode waves (but not a solution of CKP equation). As a matter of fact, if $k_1 \gg k_2, k_2 \to 0$, the big fractional expression approaches to $u_1 = \frac{k_1^2}{2t}\text{sech}^2 \frac{1}{2}(\eta_1 + \ln 3)$ which is the decay mode wave with parameter $k_1$; if $k_2 \gg k_1, k_1 \to 0$; the big fractional expression approaches to $u_2 = \frac{k_2^2}{2t}\text{sech}^2 \frac{1}{2}(\eta_2 + \ln 3)$ which is the decay mode wave with parameter $k_2$.

If $\alpha = 0$, and $\lambda y$ be a parameter, then (20) becomes the 2-decay mode solution of CKdV equation(2)

$$u(x,y,t) = \frac{2}{t} \frac{k_1^2 e^{\eta_1} + k_2^2 e^{\eta_2} + (k_1-k_2)^2 e^{\eta_1+\eta_2} + a_{12}\left(k_2^2 e^{2\eta_1+\eta_2} + k_1^2 e^{\eta_1+2\eta_2}\right)}{\left[1 + e^{\eta_1} + e^{\eta_2} + a_{12} e^{\eta_1+\eta_2}\right]^2} + \frac{x+\lambda y}{12t}$$

where $\eta_i = \frac{k_i}{\sqrt{t}}(x+\lambda y) + \frac{2k_i^3}{\sqrt{t}}, i=1,2.$

## 5. Conclusion

In this paper, we use the simplified homogeneous balance method to derive out a nonlinear transformation that from the solution of a homogemeity equation to the solution of the CKP equation. The homogemeity equation admits exponential type solutions, substituting it into the nonlinear transformation, we have the 1-decay mode and 2-decay mode solutions of the CKP equation. The 1-decay mode and 2-decay mode solutions of the CKdV equation that is the particular case of the CKP equation are also obtained. The results obtained in this paper are different from those obtained in previous literature.

**Acknoledgement**
The authors express their sincere thanks to the referee for valuable suggestions. This work were supported in part by the Basic Science and the Front Technology Research Foundation of Henan Province of China (Grant no. 092300410179) and the Scientific Research Innovation Ability Cultivation Foundation of Henan University of Science and Technology (Grant no.011CX011).